# PCIe Hot Plug support standardization challenges in ATCA

Miguel Correia[1], Jorge Sousa[1], António P. Rodrigues[1], Paulo F. Carvalho[1], Bruno Santos[1], Álvaro Combo[1], Carlos M. B. A. Correia[2], Bruno Gonçalves[1]

*Abstract*— Throughout the last decade, the Advanced Telecommunications Computing Architecture (ATCA) solidified its position as one of the main switched-based crate standards for advanced Physics instrumentation, offering not only highly performant characteristics in data throughput, channel density or power supply/dissipation capabilities, but also special features for high availability (HA), required for latest and upcoming large-scale endeavours, as is the case of ITER. Hot Swap is one of the main HA features in ATCA, allowing for Boards to be replaced in a crate (Shelf), without powering off the whole system. Platforms using the Peripheral Component Interconnect Express (PCIe) protocol on the Fabric Interface must be complemented, at the software level, with the PCIe Hot Plug native feature, currently not specified for the ATCA form-factor. From a customised Hot Plug support implementation for ATCA Node Boards, the paper presents an implementation extension for Hub Boards, allowing Hot Plug of PCIe switching devices, without causing bus enumeration problems. This paper further addresses the main issues concerning an eventual standardization of PCIe Hot Plug support in ATCA, such as the implementability of Hot Plug Elements and the generation and management of Hot Plug Events, aiming to stimulate the discussion within the PICMG community towards a long overdue standardized solution for Hot Plug in ATCA.

## I. INTRODUCTION

ADVANCED large-scale Physics experiments, such as experimental fusion devices, require hard real-time control and data acquisition systems, particularly in critical diagnostics. This is the case of the vertical stabilization (VS) controller for the Joint European Torus (JET) tokamak, developed by the Instituto de Plasmas e Fusão Nuclear (IPFN) [1]. For ITER [2], the largest fusion experiment to be built, currently under construction, in Cadarache, France, vertical control of the plasma is extremely critical, as plasma disruptions caused by loss of plasma control may severely damage the tokamak infrastructure, additionally posing both a safety and investment risk. Therefore, high availability (HA) is a key requirement for vertical stabilization control, as well as other critical diagnostics [3]. IPFN has undertaken a Research Programme on High-availability Control and Data Acquisition Systems [4], which is simultaneously contributing for the ITER Plant Control Design Handbook (PCDH) effort of standardization in fast plant system controllers (FPSC) [5], whose aim is to choose the adequate technologies and define specifications to fulfil the specific needs of fast plasma control and data acquisition at ITER.

The prototype developed by IPFN consists of hardware modules based on the PICMG 3.0 Advanced Telecommunications Computing Architecture (ATCA) [6], an industry switched-fabric modular standard which enables systems to be designed with HA, due to its robust hardware management capabilities and several types of redundancy resources, which may be used to implement fault-tolerance mechanisms. Hot Swap is one its most important features, allowing for Boards and other intelligent Field Replaceable Units (FRUs) to be replaced in a crate (Shelf), without powering off the whole system, benefiting system maintainability, thus availability.

Having ATCA been originally conceived for the telecom market, most of its commercially available products use Ethernet for communications protocol. Subsequent interest in adopting ATCA for instrumentation purposes led to the development of a specification extension for the Peripheral Component Interconnect Express (PCIe) protocol [7], a consensual choice within the instrumentation community, in particular for Physics experiments.

PCIe is a high-speed serial computer expansion bus standard [8], also providing desirable features for HA systems, such as Power Management, Quality of Service, Data Integrity, Error Handling and a Hot Plug feature for add/removal of adapter cards while keeping the system running. For form-factors other than PCIe itself, Hot Plug is implementation-dependent and should be defined by the form-factor. However, ATCA has not, yet, specified such procedures. In order to obtain complete, seamless Board insertion/extraction, this mechanism must be designed for ATCA, which constitutes a major challenge for the developed system. This paper briefly introduces the IPFN FPSC architecture, taking from an initial customized Hot Plug support implementation, developed for ATCA Node Boards, and extending it for the use case of Hub Boards, allowing the Hot

Manuscript received June 22, 2018. This work received financial support from "Fundação para a Ciência e Tecnologia" (FCT) through project UID/FIS/50010/2013.

M. Correia, J. Sousa, A. P. Rodrigues, P.F. Carvalho, A. Combo, B. Santos and B. Gonçalves are with Instituto de Plasmas e Fusão Nuclear, Instituto Superior Técnico, Universidade de Lisboa, 1049- 001 Lisboa, Portugal (e-mail:miguelfc@ipfn.tecnico.ulisboa.pt).

C.M.B.A. Correia is with LIBPhys-UC, Department of Physics, University of Coimbra, P-3004 516 Coimbra, Portugal.

Plug of PCIe switching devices, previously known to cause bus enumeration problems [9]. The presented work further addresses the main issues concerning an eventual standardization of PCIe Hot Plug support in ATCA, suggesting an architectural strategy towards its realization, with the intent of stimulating the discussion within the PICMG community.

## II. IPFN FPSC Architecture

The developed fast controller architecture consists of one ATCA crate (Shelf), with capacity for 14 Boards – Front Boards and respective Rear Transition Modules (RTMs), populated with the following types of IPFN-developed hardware modules: (i) ATCA-IO-Processor [10] – Digitizing (Node) Boards with data processing capabilities (48 analog IO channels, 18-bit ADC, 2MSPS and/or 16-bit DAC, 1MSPS); (ii) ATCA-PTSW-AMC4 [11] – PCIe switch (Hub) Boards, for fast switching of ATCA-IO-Processor generated data, with external PCIe cable interface in the RTM; (iii) PCIe external Host computer with external cabling interface (commercially available) [12]. The resulting PCIe topology is star (one switch and several endpoints). Dual-star can be implemented by using two Hub Boards with replicated Node Boards, thus enabling 2N redundancy [13].

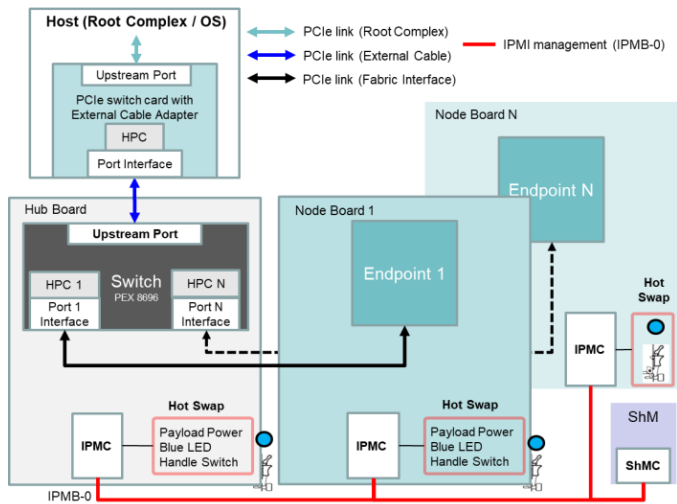

Fig. 1. Architecture of the IPFN Fast Plant System Controller ATCA platform, denoting the Hardware Platform Management system (bottom), composed by the ShMC and IPMCs of each Node and Hub Board; and the established PCIe network in star topology (top), composed by the Node Board endpoints, linking through the Fabric Interface to the PEX 8696 switch downstream ports, connected via External Cable interface to the Host.

Fig. 1 shows the resulting architecture for one Hub (Board) and N Nodes (Boards). On the bottom, the ATCA Hardware Platform Management (HPM) system is managed by the Shelf Manager (ShM); its Shelf Manager Controller (ShMC) is responsible for monitoring Board health status, by communicating with respective Intelligent Platform Management Controller (IPMC), via the Intelligent Platform Management Interface (IPMI) protocol [14], through the Intelligent Platform Management Bus (IPMB-0). The HPM system also manages the Hot Swap procedures, which allow to selectively insert/extract Boards without having to power off the Shelf. This is accomplished through specified sequences of FRU states, which are initiated by a human operator, using the Board Handles to trigger Board Activation/Deactivation requests from the Board IPMC to the ShMC [15]. Once these requests are accepted, the operator is notified by a signalling LED that Board is activated, or that it has been deactivated and may be safely extracted.

The top portion of Fig. 1 contains the resulting PCIe network. Data from each Node endpoint is handed to its respective downstream port of the PCIe switch device in the Hub Board and is sent upstream, to the Host, via External Cable. PCIe switching is accomplished by way of a Broadcom PEX 8696 PCIe Gen 2 switch [16]; This programmable device switches up to 96 Lanes, configured in 4-Lane width downstream ports for the Fabric Interface links, and a 16-Lane upstream port. PCIe Hot-Plug is supported for all downstream ports, which contain respective Hot Plug controllers, establishing a distributed control architecture, able to generate Hot Plug events, sent as software interrupts to the Host, allowing graceful add/removal of endpoint device driver and safely open/close software applications using the endpoint.

## III. Customised Hot Plug Support

As mentioned previously, one of the major challenges was to implement Hot Plug support for the FPSC platform, as it is not specified by ATCA. For the Hot Plug events to be generated, the PCIe Base Specification defines a set of Hot Plug (hardware) elements (e.g. Attention Button, Manually-operated Retention Latch, Power Controller, Attention and Power Indicators) to be connected to each HPC Port Interface; these elements do not exist in the ATCA specification and therefore are not provided in the developed FPSC Boards.

### A. Hot Plug of Node Boards

A customized solution has been initially implemented for Hot Plug of Node Boards [17], using the Node IPMC to additionally message the Hub IPMC, during the standard Hot Swap procedure, reporting that a Hot Plug operation must be initiated. The Hub IPMC then writes directly in the HPC of the corresponding Node Board to generate the necessary Hot Plug event. Finally, the HPC normally issues the corresponding Hot Plug interrupt, which is sent in-band, upstream through the External Cable, to the PCIe Host. This implementation has been made possible, since the PEX 8696 provides a test mode which allows HPC registers to be software written (via local $I^2C$ bus). Fig. 2 highlights (in yellow) the remote location of Node 1 HPC in respect to its actual endpoint device, showing the additional messaging procedures between Node 1 IPMC, Hub IPMC and the Node 1 HPC, implemented to generate the required Hot Plug events at the PEX 8696 switch.

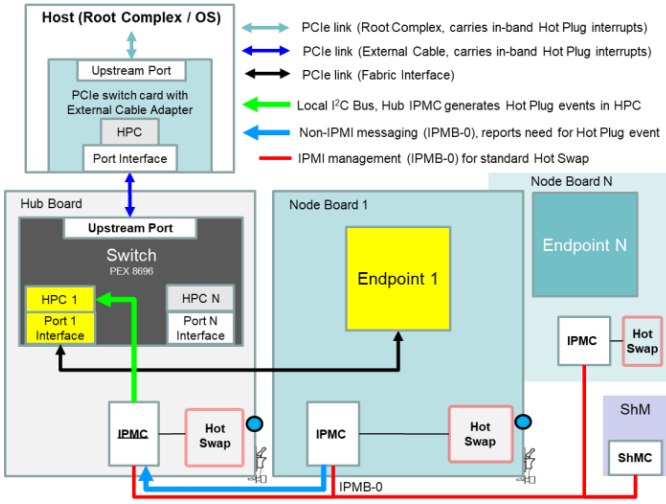

Fig. 2. Customized implementation of Hot Plug for ATCA Node Boards using IPMC messaging (along with the standard Hot Swap procedure) to trigger Hot Plug events and send respective interrupts upstream.

### B. Hot Plug of Hub Boards

Hot Plug of PCIe switch devices is known to cause PCIe bus enumeration issues which prevent new devices of being added – an issue not exclusive to ATCA implementations, since it is acknowledged by PCI-SIG [9]. In fact, the distributed HPC architecture shows that the HPC of the PEX 8696 switch is located upstream, outside the ATCA Shelf, at the downstream port of the Host External Cable adapter. This means that Hot Plug of the Hub Board must follow the Hot Plug definitions for this PCIe hardware device, as per the Standard Hot Plug Controller specification [18]. Still, it is the ATCA Hub Board who must trigger the necessary Hot Plug event, namely by toggling the Cable Present (CPRSNT#) sideband signal. In analogy with the previous procedure for Node Boards, this was achieved by additionally controlling the state of CPRSNT#, using a general-purpose input/output (GPIO) port, again in coordination with the standard ATCA Hot Swap sequences (Activation or Deactivation), as shown in fig. 3.

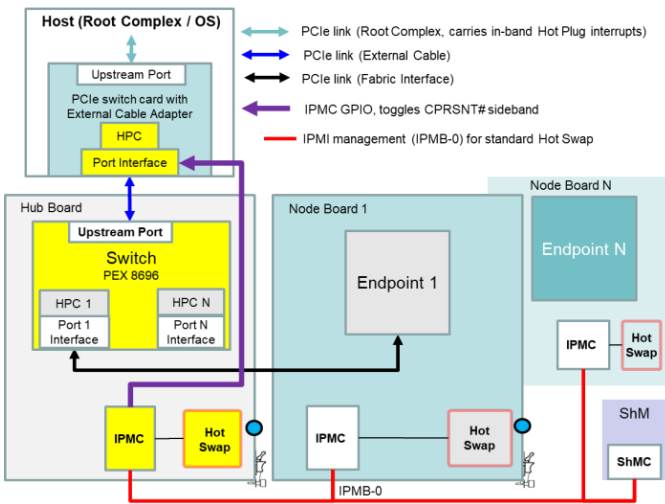

Fig. 3. Customized implementation of Hot Plug for ATCA Hub Boards, using the Hub IPMC to control the External Cable CPSRNT# sideband signal to generate the specified Hot Plug events at the respective HPC, located at the PCIe External Cable adapter board, at the Host computer.

### C. Software and Operating System

Initial developments for Hot Plug support implementation for the IPFN FPSC were performed using Redhat Enterprise Linux operating system (OS), as per ITER PCDH [19]. With this setup, Hot Plug of the PEX 8696 device (i.e. Hub Board) led to occasional OS instabilities due to the aforementioned bus enumeration issues. It was decided to use another Linux-based release – Fedora 27 [20], which supports newer kernel versions that include additional boot configuration parameters concerning Hot Plug operations [21]. Several configurations were tested, in a trial-and-error fashion, to understand how these parameters affected PCIe bus enumeration, on OS boot and upon Hot Plug operations. By setting the *hpbussize* parameter to a specific value (depending on other PCIe devices existing in the global PCIe hierarchy), successful Hot add and removal of the PEX 8696 was achieved, reflected on selected devices appearing (or disappearing) from the device list (*lspci* OS prompt command). Not only the PEX 8696 switch device has been successfully hot-plugged, but also endpoints connected to the switch (Node Board endpoints) were automatically hot-plugged, leading to seamless operation of the developed test-bench application, monitoring data from the ATCA-IO-Processor Boards [22].

## IV. TOWARDS HOT PLUG SUPPORT STANDARDIZATION

As explained in the preceding sections, ATCA Hot Swap and PCIe Hot Plug coexist as separate procedures, each managed by its own management system – Hot Swap is managed by the HPM system, whereas Hot Plug is managed by the distributed HPC architecture. This means that Board physical insertion/extraction (hardware level) is managed by ATCA Hot Swap, while PCIe Hot Plug procedures perform device add/removal at software level only. Therefore, PCIe Hot Plug add/removal event sequence must be coordinated with ATCA Hot Swap Activation/Deactivation sequences; however, neither of each of the management systems is aware of the other, which can lead to system failure, if coordination between Hot Swap and Hot Plug is lost (e.g. OS hang-up due to software bug). Furthermore, the customised solution uses non-IPMI messaging on the IPMB-0 bus, which constitutes a violation to the ATCA specification. A future Hot Plug standardization for ATCA should use the ShMC to additionally perform Hot Plug messaging, as only the ShMC is aware of all Board FRU states in the Shelf and would then be able to perform and verify coordination between Hot Swap and Hot Plug, using standard IPMI messaging on IPMB-0. In this scenario, a future ATCA specification extension for Hot Plug support should additionally: (i) Add new FRU definition fields for Hot Plug capabilities of Boards, allowing the ShMC to identify hot-pluggable devices; (ii) Define which are the required Hot Plug (hardware) elements needed in an ATCA Board, and their connection paths to respective HPCs; (iii) Contemplate software triggering of events, for which the PCIe switching hardware must be capable (PEX 8696 supports this feature but other devices may not support it); (iv) Contemplate the Hot Plug capabilities of upstream hardware and software (e.g. external cabling sideband signals connecting to Host), for Hot Plug of Hub Boards.

## V. Summary

The Hot Plug feature has generally been of limited adoption by PCI-SIG, and, consequently, poorly supported by instrumentation standards. However, Hot Plug support is vital to implement fault-tolerance mechanisms in PCIe-based ATCA platforms targeting HA.

As ATCA currently does not specify Hot Plug support, a customized solution for Node Boards has been developed, using additional messaging between IPMCs and software-generated Hot Plug events, coordinated with Hot Swap.

Hot Plug of Hub (switch) Boards was also successfully achieved, on specific OS environment, using additional IPMC control of External Cable sideband signals connecting to the Host computer. These solutions have been tested, showing that seamless continuity of operation at both hardware and software levels had been achieved, thus benefiting system availability.

For the desirable specification extension of Hot Plug support in ATCA, a set of guidelines has been suggested. The hypothesis implied that it would be ShMC responsible for handling Hot Plug messaging, requiring new definitions for FRU information fields and features to be added to ShMC and IPMC specifications. The development of such solution requires, therefore, the indispensable involvement from both hardware and software manufacturers (especially ShM and IPMC developers), but also closely connected OS developers.

Only a standardized Hot Plug solution will drive the industry to deliver compliant products (e.g., switches, FPGA PCIe cores, OSs). In this respect, the PICMG xTCA for Physics Committee (including IPFN), has taken a first step, by releasing a Hot Plug Design Guide for MicroTCA [23].

The work carried out aims to further encourage the involved communities in the efforts of standardization – towards a specification extension for the PICMG xTCA standards, to cater for the HA requirements in upcoming large Physics experiments.